\documentclass[preprint]{aastex}
\usepackage{emulateapj5}
 
\def\HI{H~{\sc i}} 
\def\HII{H~{\sc ii}}
\def\kms{${\rm km~s^{-1}}$}

\shortauthors{MCCLURE-GRIFFITHS ET AL} 
\shorttitle{THE SGPS: THE TEST REGION}
\slugcomment{Accepted by the Astrophysical Journal}
\begin{document} 

\title{The Southern Galactic Plane Survey: The Test Region}
\author{N. M. McClure-Griffiths,\altaffilmark{1}
A. J. Green,\altaffilmark{2} John M. Dickey,\altaffilmark{1}
B. M. Gaensler,\altaffilmark{3,5} R. F. Haynes,\altaffilmark{4} \&
M. H. Wieringa\altaffilmark{4}}

\altaffiltext{1}{Department of Astronomy, University of Minnesota, 116
  Church Street SE, Minneapolis, MN 55455; \mbox naomi@astro.umn.edu,
  \mbox john@astro.umn.edu}

\altaffiltext{2}{Astrophysics Department, School of Physics, Sydney
  University, NSW 2006, Australia; \mbox agreen@physics.usyd.edu.au}

\altaffiltext{3}{Center for Space Research, Massachusetts Institute of
  Technology, 70 Vassar Street, Cambridge, MA 02139; \mbox bmg@space.mit.edu}

\altaffiltext{4}{Australia Telescope National Facility, CSIRO, P.O. Box 76,
  Epping, NSW 2121, Australia; \mbox rhaynes@atnf.csiro.au, \mbox
  mwiering@atnf.csiro.au}

\altaffiltext{5}{Hubble Fellow}

\authoraddr{Address correspondence regarding this manuscript to: 
                N. M. McClure-Griffiths
                Department of Astronomy
                University of Minnesota
                116 Church St. S.E.
                Minneapolis, MN 55455}
\begin{abstract}
  The Southern Galactic Plane Survey (SGPS) is a project to image the \HI\ 
  line emission and 1.4~GHz continuum in the fourth quadrant of the Milky
  Way at high resolution using the Australia Telescope Compact Array (ATCA)
  and the Parkes Radio Telescope.  In this paper we describe the survey
  details and goals, present $\lambda 21$-cm continuum data, and discuss
  \HI\ absorption and emission characteristics of the SGPS Test Region
  ($325\fdg5 \leq l \leq 333\fdg5$; $-0\fdg5 \leq b \leq +3\fdg5$).  We
  explore the effects of massive stars on the interstellar medium (ISM)
  through a study of \HI\ shells and the \HI\ environments of \HII\ regions
  and supernova remnants.  We find an \HI\ shell surrounding the \HII\ 
  region RCW~94 which indicates that the region is embedded in a molecular
  cloud.  We give lower limits for the kinematic distances to SNRs
  G327.4+0.4 and G330.2+1.0 of 4.3~kpc and 4.9~kpc, respectively.  We find
  evidence of interaction with the surrounding \HI\ for both of these
  remnants.  We also present images of a possible new SNR G328.6-0.0.
  Additionally, we have discovered two small \HI\ shells with no
  counterparts in continuum emission.
\end{abstract}

\section{Introduction}
\label{sec:intro}
Surveys of the Milky Way allow us to look at the energetics and structure of
our own Galaxy with spatial resolution that is unattainable in other
galaxies.  The inner Galaxy, in particular, provides a number of scientific
opportunities.  In the 1960's and 70's Galactic neutral hydrogen (\HI) and
continuum were mapped using single dish radio telescopes with low spatial
resolution (eg.\ Weaver \& Williams 1974\nocite{weaver74}; Goss \& Shaver
1970\nocite{goss70}).  As a result, much of what we know about the structure
of the Galactic \HI\ is restricted to large scales. In other wave-bands
Galactic surveys are much more up-to-date.  In the infrared, X-ray, and
H$\alpha$ the inner Galaxy has been mapped extensively, while \HI\ remains
seriously under-sampled.  Over the past five years the \HI\ atlas of the
Galaxy has been greatly improved by the Dominion Radio Astrophysical
Observatory's (DRAO) Canadian Galactic Plane Survey (CGPS) which covers a
longitude range of $l= 74$\arcdeg\ to 147\arcdeg\ using a combination of
interferometer and single dish data to image the Galaxy at a resolution of
one arcminute \citep{taylor99}.  Despite the contributions of the CGPS to
our knowledge of Galactic \HI, the inner Galaxy remains neglected.

We have recently begun the Southern Galactic Plane Survey (SGPS), a
large-scale project to image the $\lambda 21$-cm continuum and \HI\ spectral
line in the fourth quadrant of the Galactic Plane with high angular and
velocity resolution \citep{dickey99,mcgriff99}.  The SGPS makes use of high
spatial resolution data from the Australia Telescope Compact Array (ATCA)
and short spacing information from the Parkes 64m single dish.\footnote{The
Parkes telescope and the ATCA are part of the Australia Telescope, which is
funded by the Commonwealth of Australia for operation as a national facility
managed by the Commonwealth Scientific and Industrial Research
Organisation.}  The final project will provide a complete \HI\ dataset of
$253\arcdeg \leq l \leq 358$\arcdeg\ and $-1\fdg0 \leq b \leq +1\fdg0$ at an
angular resolution of 2\arcmin, and with velocity resolution of $\Delta v =
0.82~{\rm km~s^{-1}}$.  In addition, we have extended the single dish
coverage to $b=\pm 10\arcdeg$ in order to study large scale structures which
protrude from the Galactic plane.

This dataset is particularly useful for studying the structure and dynamics
of the neutral medium, on which massive stars have a significant impact.  O
and B stars not only affect the medium through ionization, as in the case of
\HII\ regions, they also perturb the medium through winds and at the end of
their lives, as supernovae.  The resultant supernova remnants (SNRs) plow
shocks through the neutral medium, ionizing and compressing the medium and
leaving a lasting impression.  Not only do \HII\ regions and SNRs impact the
ISM, the structure of the ISM - particularly density enhancements - affects
the morphology of \HII\ regions and SNRs.  After the SNR or \HII\ region has
ceased to exist in continuum, the imprint may remain in the ISM in the form
of an \HI\ shell.  Because of the large range of spatial scales sampled with
the combined Parkes and ATCA data, the SGPS is an ideal dataset in which to
explore these effects.

In this paper we introduce the details of the Southern Galactic Plane Survey
with attention to \HI\ spectral line and $\lambda$21-cm continuum data from
the SGPS Test Region ($325\fdg5 \leq l \leq 333\fdg5$; $-0\fdg5 \leq b \leq
+3\fdg5$).  Other scientific highlights from the SGPS are discussed
elsewhere: two large \HI\ shells discovered in the Parkes data are presented
in \citet{mcgriff00b}, preliminary images of \HI\ emission and absorption
features are presented in \citet{mcgriff00a}, and the polarization
properties of the Test Region are presented in \citet{gaensler00a}.  Here we
explore the connections between the \HI\ and $\lambda$21-cm continuum images
of the Test Region.  In \S\ref{subsec:obj} \& \ref{sec:obs} we describe the
survey objectives, observing and data analysis strategies.  In
\S\ref{sec:cont} we discuss the $\lambda$21-cm continuum emission.  \HI\
absorption towards continuum sources is discussed in \S\ref{sec:abs}.  We
have chosen a representative sample of \HII\ regions and supernova remnants
(SNRs) to study the relationship between the continuum emission from these
objects and the surrounding \HI\ environments in \S\ref{sec:emis}.  The Test
Region is an excellent area to initiate such a study as it contains many
\HII\ regions and SNRs, as well as extended emission structure.  Using \HI\
absorption and \HI\ morphological matches to the continuum emission, we seek
to create a three-dimensional view of the Galaxy in this subregion.

\subsection{Survey Objectives}
\label{subsec:obj}
The general goal of the Southern Galactic Plane Survey is to provide a
dataset with which to study the structure and dynamics of the neutral
hydrogen (\HI) in the inner Galaxy.  Previous studies of the inner Galaxy
have lacked the sensitivity and resolution necessary to study the physical
processes of the interstellar medium (ISM) over a large range of spatial
scales.  Though the specific goals of the SGPS are numerous, we will
highlight a few below:
\begin{itemize}
\item The SGPS will allow us to address questions about the spatial
distributions and scale heights of the warm and cool components of the
neutral medium in order to understand the thermal phases of the interstellar
medium.

\item Combining the Parkes and ATCA data, we will be able to probe the
interstellar medium through a broad range of spatial scales in order to
develop a statistical interpretation of the neutral ISM throughout the inner
Galaxy.

\item \HI\ emission data over the large range of spatial scales available in
the SGPS will allow us to detect a full sample of \HI\ shells with which to
study not only the formation of shells - particularly of the largest, most
enigmatic ones - but also their distribution in the Galaxy and global effect
on Galactic structure.

\item A number of \HI\ self-absorption (HISA) features are apparent in the
SGPS Test Region \citep{mcgriff00a}.  HISA, where cold \HI\ clouds absorb
the diffuse background \HI\ emission, is an excellent probe of the
distribution of the coldest, most compact \HI\ clouds.  Further exploration
of these features will be presented in a subsequent paper \citep{dickey00}.

\item The dataset will provide \HI\ absorption spectra for distance
estimates of many Galactic objects and may be useful in identifying
those extragalactic sources located close to the Galactic plane.

\item The inclusion of full polarization information for the continuum data
will allow us to explore the polarization structure of individual objects,
such as supernova remnants, and to investigate the Galactic magnetic field
structure using the polarization of the diffuse background emission. Results
of the polarization properties of the Test Region appear in
\citet{gaensler00a}.
\end{itemize}

\section{Observations and Analysis}
\label{sec:obs}
Observations of the SGPS Test Region were made with the Australia Telescope
Compact Array (ATCA; Frater, Brooks, \& Whiteoak 1992) and Parkes Radio
Telescope.  The ATCA is an east-west synthesis instrument near Narrabri NSW,
with six 22~m antennas on a 6~km track.  Five antennas are movable into
configurations with baselines between 31~m and 6~km.  The ATCA data consist
of a 190 pointing mosaic covering $325\fdg5 \leq l \leq 333\fdg5$ and
$-0\fdg5 \leq b \leq +3\fdg5$.  These data were obtained during five
separate observing sessions between April 1997 and April 1998.  The
observing dates and times are given in Table~\ref{tab:obs}.  The
observations were made with several compact array configurations - 750A,
750C, 750D, and 375 - in order to obtain maximum sensitivity to large scale
structures.  Each of the 190 pointings was observed in forty 30~s snapshots
at a broad range of hour angles for good {\em u-v} coverage.  The pointings
for the Test Region were arranged on a square grid with 15\arcmin\
separation (Nyquist) as determined by the ATCA primary beam FWHM, which is
33\arcmin\ at $\lambda$-21 cm.  The pointing centers are plotted on the
21-cm continuum ATCA image of the SGPS Test Region shown in
Fig.~\ref{fig:centers}.  The ATCA feeds receive two orthogonal linear
polarizations, $X$ and $Y$. All observations were recorded in a wideband
continuum mode with 32 channels, each 4~MHz, across a 128 MHz total
bandwidth with polarization products $XX$, $YY$, $XY$, and $YX$ to enable
calculation of all four Stokes parameters; and simultaneously in a spectral
line mode with polarization products $XX$ and $YY$ in 1024 channels across a
4 MHz total bandwidth.  The continuum data are centered on $\nu = 1384$ MHz,
whereas the spectral line data are centered on $\nu= 1420.0$ MHz and have
channel separation of $\Delta v = 0.82~{\rm km~s^{-1}}$ ($3.9$ kHz).

Data editing, calibration, and imaging of the ATCA data are carried out in
the {\sc miriad} data reduction package \citep{sault00}.  The source PKS
B1934-638 is used for flux density and bandpass calibration and was observed
once per observing session.  A flux density of 14.94 Jy at 1420 MHz is
assumed \citep{reynolds94}.  PKS B1934-638 has no detectable linear
polarization and can therefore be used to solve for the polarization
leakages.  Polarization leakages are also calculated for the sources MRC
B1613-586 and MRC B1431-48 which were observed approximately once every hour
for calibration of the time-variation in complex antenna gains.

The individual pointings were linearly combined and imaged using a standard
grid-and-FFT scheme with super-uniform weighting.  Like uniform weighting,
super-uniform weighting minimizes sidelobe levels to improve the dynamic
range and sensitivity to extended structure.  Uniform weighting reduces to
natural weighting, however, if the field of view is much larger than the
primary beam, as is the case for large mosaics.  Super-uniform weighting
overcomes this limitation by decoupling the weighting from the field size
\citep{sault00}.  In this manner super-uniform weighting attempts to
minimize sidelobe contributions from strong sources over a region smaller
than the full field being imaged and therefore is typically more successful
than uniform weighting on large mosaics \citep{sault96}.  

The specifics of the calibration and imaging of ATCA polarization data are
described in \citep{gaensler00a}, here we discuss only the Stokes I
continuum and \HI\ data.  Two \HI\ data cubes were produced.  In one cube
the continuum emission was subtracted from the \HI\ {\em u-v} data using the
{\sc miriad} task uvlin.  The second cube, for use in absorption studies,
contains continuum emission.  The resultant synthesized beam for both cubes
and the continuum image is $124\farcs9 \times 107\farcs5$ ($\alpha \times
\delta$).  For the data presented here the 6~km baselines of the ATCA were
not used.  However, the long baselines are retained for some absorption
studies, such as those presented in \citet{dickey00}.  Joint deconvolution
was performed on the full linear mosaics using a maximum entropy algorithm
implemented in {\sc miriad} task mosmem \citep{sault96}. The method of joint
deconvolution is very effective for maximizing the {\em u-v} coverage
attainable in mosaiced observations \citep{sault96}.  Despite deconvolution,
some sidelobes are visible around strong point sources in the continuum and
\HI\ images.

Though mosaicing allows us to recover angular scales larger than normal
interferometric observations by reducing the effective shortest projected
baseline, we are nonetheless limited in practice to angular scales smaller
than $\theta = \lambda / (d -D/2)$, where $d=30.6$~m is the shortest
baseline of the ATCA, and $D= 22$~m is the diameter of a single
antenna. This limits the ATCA data to angular scales smaller than $\theta
\approx 36\arcmin$.  In order to recover information on larger size scales,
the ATCA mosaic must be combined with single dish data from the Parkes Radio
Telescope.

The Parkes Radio Telescope is a 64~m antenna situated near Parkes NSW,
Australia.  It has a thirteen beam, $\lambda$21-cm receiver package at prime
focus \citep{staveley-smith96}.  The SGPS Test Region was observed on 1998
December 15-16.  The Parkes survey was subsequently expanded to $b=\pm
10\arcdeg$ for more complete coverage of large scale structures during
additional observing sessions: 1998 June 18-21, 1999 September 18-27, and
2000 March 10-15.  Observations were made by the process of mapping
``on-the-fly'' with the inner seven beams of the multibeam system.  Data
were recorded in 5~s samples, while scanning through three degrees in
Galactic latitude.  The data were taken in frequency switching mode using
the narrowband back-end \citep{haynes98}, with a total bandwidth of 8 MHz
across 2048 channels.  Though the Parkes data are frequency switched, total
power information for each sample is retained.  Each sample was divided by
the previous frequency switched sample and the residual bandpass shape
fitted with a series of Fourier components.  The spectra were then
multiplied by the mean of the reference signal over the spectrum to
reconstruct the continuum emission with a flat baseline.  Absolute
brightness temperature calibration of the \HI\ line data was performed from
observations of the IAU standard regions S6 and S9 \citep{williams73}.  A
detailed description of the observing strategy, calibration, and imaging
procedures is found in \citet{mcgriff00b}.  Off-line channels were used for
continuum subtraction and to produce the continuum image using the AIPS task
IMLIN.  The Parkes data on the SGPS Test Region have an angular resolution
of $\sim 15\arcmin$.  The final, calibrated data have a bandwidth of 4 MHz
with 1024 channels, such that the velocity resolution of $0.82$~\kms\
matches the ATCA data.

It should be noted that the data presented here have not been corrected for
the effects of stray radiation.  Stray radiation leakage from bright \HI\
emission through the back sidelobes of a single dish beam introduces
baseline errors which are typically between $0.5$~K and $2$~K
\citep{kalberla80}.  When compared to low latitude Galactic spectra, this is
a small percentage, but it does nonetheless limit the sensitivity and
accurate representation of extended features in the Parkes data.  The data
for the entire survey will have a first order stray radiation correction
applied.  A complete correction requires a low resolution survey of the
entire sky with a known antenna pattern in order to reconstruct the stray
radiation at every point on the sky, at any azimuth and elevation, and at
any time of the year.  Such a task is beyond the scope of this project.  A
first order correction, however, can be done by re-observing the survey
region at different times of the year and comparing the spectra.  The
velocity shifts caused by the Earth's motion around the Sun result in excess
emission at different velocities. We have, therefore, re-observed the full
survey region four times at three month intervals around the year and we
will compare each spectrum.  The minimum value at each velocity will be a
reasonable upper limit to the stray radiation corrected profile, good to
within $\sim 0.5$~K of $T_B$.

The final step of imaging involves combining the Parkes and ATCA data.  The
data may be combined in the Fourier domain after deconvolution of the
individual images or in the {\em u-v} plane prior to deconvolution.
\citet{stanimirovic99} showed that the results are comparable using either
method, but that combining after deconvolution produced results that were
typically more consistent than with other methods.  Comparison of our data
combined in both ways shows similar results.  We have chosen, therefore, to
combine the data in the Fourier domain after deconvolution.  In this method,
the interferometric data \HI\ and continuum data are imaged and deconvolved,
the single-dish data are imaged and the clean interferometric and
single-dish images are Fourier transformed and combined.  This technique is
implemented in the {\sc miriad} task immerge.  Slight differences in
calibration can lead to the necessity of a relative calibration factor by
which the single-dish dataset is multiplied before combination.  This
calibration factor is determined by comparing the datasets in the Fourier
plane at every pixel and frequency in the range of overlapping spatial
frequencies.  In order to calculate the calibration factor both images must
be deconvolved, a step which requires a good knowledge of the single-dish
beam \citep{stanimirovic99}.  Using a two dimensional Gaussian with FWHM
$15\farcm5$ for the Parkes beam and by comparing the Parkes and ATCA
continuum images of a strong, compact source in the Test Region, we
calculated a relative calibration factor of 1.19.  Two combined \HI\ data
cubes were created, one containing the continuum emission for absorption
studies and one which had the continuum subtracted for emission studies.
The continuum images were combined in the same way as the cube and with the
same calibration factor.  The combined Parkes and ATCA 21-cm continuum image
is shown in Fig.~\ref{fig:21cont}.

The combined data are sensitive to all angular scales from the synthesized
beam size, $124\farcs9 \times 107\farcs5$ ($\alpha \times \delta$), up to
the image size, $8\arcdeg \times 4 \arcdeg$ ($l \times b$) for the Test
Region.  Because of the fine scale structure seen in the velocity domain, no
Hanning smoothing was applied to the data.  Each channel image has a
velocity separation of $0.82$~\kms.  Channel images from the continuum
subtracted combined data cube are shown in Fig.~\ref{fig:chans1}.  Every
fourth channel from $v=-127$~\kms\ to $v=79$~\kms is shown.  The rms noise
in the channel images is $\sim 2.4$ K of ${\rm T_B}$ for the ATCA data,
$\sim 100$ mK of ${\rm T_B}$ for the Parkes data and $\sim 2.3$ K of ${\rm
T_B}$ for the combined dataset.  The rms noise in the continuum images is
$\sim 5.5~{\rm mJy~beam^{-1}}$ for the ATCA data, $\sim 500~{\rm
mJy~beam^{-1}}$ for the Parkes data (beam size $15\farcm5 \times
15\farcm5$), and $\sim 7~{\rm mJy~beam^{-1}}$ for the combined data.

\section{Continuum Emission}
\label{sec:cont}
The combined $\lambda$21-cm continuum image of the SGPS Test Region is shown
in Fig.~\ref{fig:21cont}.  Most sources have been previously catalogued as
\HII\ regions or SNRs \citep{avedisova97,caswell87,green00,whiteoak96}.
There are also many unresolved sources scattered throughout the Test Region.
\HI\ absorption measurements towards many of these suggest that most are
extragalactic.  This region has been studied in H${\rm \alpha}$ by
\citet{georgelin94} as part of an extensive H${\rm \alpha}$ survey of the
Southern Galactic Plane.  As shown in the Fig.~\ref{fig:diag}, a diagram of
the expected velocities and spiral arms in the fourth quadrant, the Test
Region line of sight crosses both the Sagittarius-Carina and Scutum-Crux
spiral arms and runs tangent to the Norma arm at $l\approx 327\arcdeg$.  As
a result this region has a particularly high density of continuum sources.

\subsection{Discrete Sources}
\label{subsec:sources}
We describe here the more prominent discrete sources in the SGPS Test
Region.  These sources are marked in on the MOST 843~MHz continuum image
shown in Fig.~\ref{fig:most}.  Several individual sources are discussed in
detail below with comment given about their associated \HI\ emission.
Starting at the lower longitude end, the first strong source is RCW~94
\citep{rodgers60, shaver79} at $l=326\fdg3$, $b=+0\fdg8$, with an angular
diameter of about 18\arcmin.  This ring-like structure is an \HII\ region,
with strongest emission to the lower left.  There is a smaller region
adjoining the \HII\ region at $l=326\fdg4$, $b=+0\fdg9$.  At $l=326\fdg7$,
$b=+0\fdg8$ is another \HII\ region, RCW~95.  Directly below RCW~95 is the
brighter, extended \HII\ region G326.65+0.59 \citep{georgelin94}.  Closer to
the Plane at higher longitudes is a very extended thermal filamentary
structure G326.96+0.03.  This source has arcs of emission above and below a
centralized bright knot.  Because these sources are all at the same
distance, we refer to the grouping of RCW~94, RCW~95, G326.65+0.59, and
G326.96+0.03 as the RCW~94-95 \HII\ region complex.

Above the high longitude edge of the G326.96+0.03 arc is SNR G327.4+0.4, a
large shell type SNR with enhanced limb brightening to the lower left.
Further from the Plane than G327.4+0.4 there is a smaller, weaker supernova
remnant, SNR G327.4+1.0.  This source has a nearly closed arc extending to
higher latitudes.  At slightly higher longitudes there is a region of
extended emission comprised of several thermal sources grouped at
G327.83+0.11.  At higher longitudes and lower latitudes than these sources
there is another \HII\ region, G327.99-0.09.  Near $l=328\arcdeg$ is the
compact \HII\ region G328.31+0.45 and the extremely bright Crab-like SNR
G328.4+0.2 \citep{gaensler00b}.

At higher longitudes, the compact source G328.81-0.08 is classified as an
\HII\ region in \citet[ hereafter CH87]{caswell87} on the basis of a
recombination line detection.  However, examination of the Midcourse Space
Experiment (MSX; Egan et al.\ 1998\nocite{egan98}) band A ($6.8-10.8$
\micron) image shows only a small infrared source, IRAS 15550-5306, slightly
offset from the center of G328.81-0.08.  This infrared source has a FWHM of
$\sim 30\arcsec$, whereas the offset $\lambda$21-cm source has a FWHM of
$\sim 3\arcmin$.  It is unclear whether the infrared source is the same as
the radio source.

About $30\arcmin$ from G328.81-0.08 there is an extended source centered at
$l=328\fdg6$, $b=0\arcdeg$.  Fig.~\ref{fig:newsnr} shows the combined Parkes
and ATCA 1.4~GHz SGPS continuum image of this region and the MOST 843~MHz
image of the same area.  In the SGPS image the source has a mostly filled,
double loop morphology of angular diameter $\sim 0\fdg5$, and internal
filamentary structure.  Only the edges of the source are observed as
filaments at the $\sim 15~{\rm mJy~beam^{-1}}$ level in the MOST image.  The
smooth, extended emission observed in the SGPS image is not detected by
MOST.  There is no counterpart to this emission in the MSX band A image.
The fact that no corresponding IR emission is detected from this source
suggests that the emission is non-thermal and we propose that it is a new
SNR candidate, G328.6-0.0.  There are also several long overlapping thermal
filaments visible in both of the radio continuum images and the MSX image.
These thermal filaments do not appear to be directly associated with this
double loop SNR candidate, G328.6-0.0, but extend to $b=0\fdg5$.  Embedded
within the left edge of the SNR candidate is an unresolved source at
$l=328\fdg59$, $b=-0\fdg11$ which does not appear in the MSX image,
suggesting that it too is either non-thermal or extragalactic.

Near $l=329\arcdeg$ there are two \HII\ regions, G329.35+0.14 and
G329.49+0.21.  At larger longitudes there is the large supernova remnant,
SNR G329.7+0.4, which accounts for much of the extended emission in this
portion of the Test Region.  The large ($\Delta \theta \sim 40\arcmin$)
remnant consists of many loops.  Above G329.7+0.4 is a large ($\sim
20\arcmin$) loop of thermal emission which is also visible in the MSX
images.  This may be an extended \HII\ envelope (EHE) as similarly suggested
for the area around $l=312\arcdeg$ \citep{whiteoak94}.  This region appears
connected to large arcs of thermal emission extending above and below SNR
G330.2+1.0.  SNR G330.2+1.0 is a composite remnant with irregular emission
that does not seem to form a clear shell.  Between $l=330\fdg2$ and
$l=330\fdg9$ there are few bright discrete sources.  The area is largely
filled with extended emission.

Another large \HII\ complex is located around $l=331\arcdeg$, including the
\HII\ regions: G331.03-0.15, G331.26-0.19, G331.52-0.07; and at the high
longitude limit of the Test Region: SNR G332.0+0.2.  SNR G332.0+0.2 is a
mostly complete shell of angular diameter $\sim 10\arcmin$. Further above
the Plane are two large \HII\ regions: G331.35+1.07 and G331.36+0.51.

\subsection{Extended Continuum Emission}
\label{subsec:extended}
Throughout the Test Region there is diffuse emission which decreases in
intensity with increasing latitude.  Fig.~\ref{fig:contslice} shows three
slices across the Test Region continuum image, all at $l=328\fdg7$.  The
first slice is across the ATCA image alone, the second across the Parkes
image alone, and the third is across the combined image. It is clear that
the ATCA data is not sensitive to the large scale Galactic emission, but
resolves the individual sources.  The slice across the Parkes image shows
the large-scale decrease in emission, but the low resolution does not
clearly delineate the individual sources.  The combined image, however,
shows both the large-scale emission and the resolved sources.  From this
slice it is obvious that both single dish and interferometric data are
necessary to understand the relationship between discrete and extended
Galactic continuum emission.

Though some of the extended emission in the Test Region is smooth and can be
attributed to the diffuse Galactic background, there is also structure in
the emission.  It is not immediately obvious whether the structure can be
attributed to the Galactic background, whether it is associated with
discrete continuum sources, or whether it has an altogether different
nature.  This question was addressed by \citet{whiteoak94} who carefully
examined a $3\arcdeg \times 2\arcdeg$ region in the Plane around
$l=312\arcdeg$ from the MOST Galactic Plane Survey \citep{green99}.  They
note low surface brightness, extended emission in the high resolution
843~MHz continuum images and suggest that this is associated with the
ionized ISM.  The emission they detect is thermal and they relate some of it
to extended \HII\ envelopes (EHEs) around \HII\ regions.  The MOST data are
only sensitive to structures up to 30\arcmin, so it is unclear whether there
are larger-scale associations amongst some of these filamentary structures.
The SGPS data are sensitive to all size scales up to about four degrees and
are hence well-suited to study these features.

The MOST and SGPS surveys complement each other very well.  Comparison of
the MOST 843~MHz image in Fig.~\ref{fig:most} and the SGPS ATCA continuum
image in Fig.~\ref{fig:centers} shows good correlation between the two
interferometric images, though the MOST images are at a slightly higher
angular resolution ($43\farcs0 \times 51\farcs9, \alpha \times \delta$).  In
both images there are filaments, typically a few arcminutes in width and up
to few degrees in length.  With the inclusion of the Parkes data in
Fig.~\ref{fig:21cont} we can observe how these filaments relate to the
larger-scale diffuse emission.  It appears that there are two categories of
extended emission: one where filamentary structures are part of
larger-scale, filled structures and one where the filaments are
self-contained structures.  An example of a structure which appears
filled-in with the inclusion of the Parkes data is the possible SNR
G328.6-0.0, as shown in Fig.~\ref{fig:newsnr} and described in
\S\ref{subsec:sources}.  In this case, the loops observed in the MOST and
ATCA images are observed as part of a cohesive, filled structure in the SGPS
image.  Clearly, the filaments are the SNR edges which the interferometers
can detect, while they cannot detect the large-scale smooth emission in the
center of the remnant.  By contrast, there are many loops and filaments near
SNR G330.2+1.0 that appear filamentary in both the MOST and SGPS images,
implying that they are not part of a larger, filled structure.  These
filaments are visible in the MSX images, implying that they are thermal.
These structures may be sheets viewed edge-on, threads, or the edges of EHEs
where the surfaces are too diffuse to be detected.
\section{\HI\ Absorption}
\label{sec:abs}
In order to create a three-dimensional view of this portion of the Galaxy we
have extracted \HI\ absorption spectra towards the brighter \HII\ regions
and SNRs.  The method we use to determine the HI absorption spectrum is
based on averaging the spectra toward the brightest part of the continuum
("on-source spectra"), and subtracting an interpolated average of the
spectra from the region surrounding the continuum source ("off-source
spectra"; Dickey et al.\ 1992\nocite{dickey92}).  The on-source spectra are
selected based on the continuum image by setting a high threshold, typically
80\% of the continuum peak, and including in the average only spectra toward
pixels whose continuum brightness is above this high threshold.  These
on-source spectra are averaged with weighting factor equal to the continuum
brightness in each pixel, which optimizes the signal-to-noise ratio in the
resulting absorption spectrum.  Similarly, the off-source spectra are
selected toward pixels whose continuum brightness is below a low threshold,
typically 20\% of the continuum peak.  The off-source spectra are not simply
averaged, but interpolated to give a better prediction of the emission
spectrum in the direction of the continuum peak.  This interpolation is
based on a simple bi-linear fit (least squares fitting a linear function of
two dimensions) done independently for each spectral channel.  The outer
boundary for which spectra are included in the off-source interpolation is
typically 7\arcmin, but for the extended supernova remnant G327.4+0.2 we
extend the outer boundary to 15\arcmin.  The high and low thresholds also
have to be adjusted in some cases depending on the continuum flux, down to
70\% for the high threshold of the faintest sources, and in the range 10\%
to 30\% for the low threshold depending on the angular size of the continuum
distribution.  This interpolation process does not change the fundamental
angular resolution of the survey, i.e.\ it is not an extrapolation on the
{\em u-v} plane.  So the spectra derived for both the absorption and the expected
emission still correspond to one beam area of roughly 2\arcmin\ diameter.
                                                                              
The absorption spectrum is then determined by subtracting the interpolated
off-source spectrum from the averaged on-source spectrum.  The resultant
optical depth spectrum, $e^{-\tau}$, is the absorption spectrum divided by
the continuum flux averaged over the pixels above the high threshold.  The
optical depth spectra towards RCW~94 and G326.65+0.59 are shown in
Fig.~\ref{fig:rcw94_abs} (bottom plot) with corresponding interpolated
off-source emission spectra (upper plot).  The errors in the absorption
spectra are generally dominated by uncertainty in the interpolated emission,
particularly at low latitudes where the emission is not smoothly distributed
on scales of a few arcminutes.  To estimate the error in the absorption we
compute the predictions for the off-source spectra based on the bilinear fit
to the emission and take the difference between these predictions and the
actual spectra in each off-source pixel.  The rms average of these
differences gives an error envelope on the interpolated emission in the
direction of the continuum peak.  Dividing by the continuum flux averaged
over the on-source pixels gives the optical depth error spectrum ($\pm 1
\sigma$), shown in dotted lines on
Figs.~\ref{fig:rcw94_abs}~\&~\ref{fig:g327abs}.

For the absorption feature with most extreme velocity (most negative or most
positive) we define $V_{\rm L}$, the lower limit on the distance.  Following
\citet{frail90} we define the velocity corresponding to the upper distance
limit, $V_{\rm U}$, as the first emission peak ($T_b > 35$~K) beyond $V_{\rm
L}$ which does not correspond to any absorption features. Upper distance
limits are estimates only, since the absence of an absorption line at the
higher velocity is not conclusive evidence that the continuum source is
nearer than the HI emission.  A region of 21-cm emission may not show
absorption because the cool gas may have a covering factor less than one
(i.e. there may be gaps between absorbing clouds, filled only with warm gas
which does not show detectable absorption).  This is unlikely for an
emission region with column density of $3 \times 10^{20}~{\rm cm^{-2}}$ or
more, which is implied by a line of brightness temperature greater than
35~K.  Emission lines stronger than this almost always show some absorption,
so our upper limit distances should be mostly valid.  We assume that the
velocity errors are dominated by random cloud motions on the order of 6~\kms
\citep{dickey97}.  Velocity limits and corresponding kinematic distances are
given in Table~\ref{tab:abs} for sources brighter than $\sim 800~{\rm
mJy~beam^{-1}}$.  Included in the table for comparison are the radio
recombination line velocities from CH87.  Some individual sources are
discussed in detail later.

The absorption velocities in Table~\ref{tab:abs} clearly lie in two dominant
distributions, one centered around $v=-50$~\kms\ and another centered around
$v=-90$~\kms.  Calculated isovelocity contours from the \citet{fich89}
rotation curve are plotted in Fig.~\ref{fig:diag} on the \citet{taylor93}
model of Galactic spiral arms to show the velocities covered by individual
spiral arms and the range of expected velocities for any given line of
sight.  The Test Region line of sight is marked by the wedge.
Fig.~\ref{fig:diag} shows that gas between $v=-20$~\kms\ and $v=-30$~\kms\
is located in the Sagittarius-Carina arm.  The position of this arm in
velocity space is well traced by \HII\ regions
\citep{georgelin76,caswell87}.  Gas at velocities between $v=-50$~\kms\ and
$v=-75$~\kms\ is located within the Scutum-Crux arm.  As noted in CH87 there
are many features around $v=-90$ \kms\ for which the correspondence to a
spiral feature is unclear.  \citet{georgelin76} assign these \HII\ regions
to the Norma arm.  At $l\approx 327\arcdeg$ the line of sight is nearly
tangent to this arm, which accounts for the large number of sources seen
there.  The distribution centered at $v=-90$~\kms\ has a large spread in
velocity space, extending as far as the terminal velocity near
$v=-110$~\kms.  This large velocity spread can also be explained by the line
of sight remaining in the spiral arm for a significant distance near the arm
tangent.  Using \HII\ regions and diffuse H$\alpha$ emission,
\citet{georgelin94} similarly note velocity distributions at $v=-20$~\kms,
$v=-40$~\kms, and $v=-65$~\kms.  They do not, however, detect H$\alpha$
emission near $v=-90$~\kms\ because it is beyond the extinction limit.

From the \HI\ absorption we determine new kinematic distances to two
supernova remnants and confirm distances to a further one SNR and nine \HII\ 
regions.  For SNR G328.4+0.2 our absorption spectrum looks very similar to
\citet{gaensler00b}, who also found an extreme velocity of $v=28$~\kms.  The
\HII\ region velocities all correspond with the velocities given in CH87.
The new kinematic distances for SNR G327.4+0.4 and SNR G330.2+1.0 are given
in Table~\ref{tab:abs} and are discussed in detail below.

\subsection{Distance Ambiguities for Individual Sources}
\label{subsec:ambi}
Many of the sources presented here have distance ambiguities.  Sources in
the fourth quadrant with negative velocities are found inside the solar
circle where each velocity corresponds to two distances.  There are several
methods for distinguishing between the two distances.  H$\alpha$ emission is
often used as an indicator, as it is severely absorbed at far distances.
One can also make rather uncertain estimates based on associations with
nearby objects of known distance, the emission spectrum towards the object,
or the more likely linear size and luminosity.  For \HII\ regions with
recombination velocities, comparison of the \HI\ absorption velocity with
the recombination line velocity can resolve the ambiguity \citep{kuchar94}.
If the most extreme \HI\ absorption is at or near the recombination line
velocity then the cloud is at the near distance.  However, if the \HI\ 
absorption is seen beyond the recombination line velocity then the \HII\ 
region is at the far distance.

For all of the \HII\ regions presented here we have recombination line
velocities from CH87.  In all cases there is no absorption significantly
beyond the recombination line velocity, implying the near distance for these
regions.  The \citet{georgelin94} H$\alpha$ survey of this region also
resolved many of the distance ambiguities through associations with stellar
distances.  In particular, \citet{georgelin94} favor the near distance for
the star forming region associated with RCW~94 and RCW~95, as discussed
below.

\section{\HI\ Emission Features}
\label{sec:emis}

\subsection{The  RCW~94-95 \HII\ Region Complex}
\label{subsec:rcw}
As described in Section~\ref{sec:cont}, the Test Region contains many
catalogued \HII\ regions.  RCW~94, RCW~95, and G326.65+0.59 are part of a
large star-forming complex in the Scutum-Crux arm.  ${\rm H_2CO}$ and
hydrogen recombination line (H$109\alpha$ \& H$110\alpha$) velocities of
$v=-42$~\kms\ and $v=-45$~\kms have been measured for RCW~94 and 95,
respectively (CH87).  We have extracted \HI\ absorption profiles towards
both RCW~94 and G326.65+0.59 (see Fig.~\ref{fig:rcw94_abs}) which confirm
these velocities, showing deep absorption out to $v=-47$~\kms\ and
$v=-43$~\kms, respectively.  The most extreme absorption line for
G326.65+0.59, centered at $-46$~\kms\, is slightly broader ($\Delta v
\approx 15$~\kms) than that for RCW~94 ($\Delta v \approx 9$~\kms),
suggesting a more turbulent region \citep{shaver79}.  \citet{georgelin94}
also note the differing line widths of these two regions.  We adopt the IAU
standard values for the Sun's orbital velocity, $\Theta_o = 220$~\kms,
Galactic center distance, $R_o = 8.5$~kpc, and use the rotation curve of
\citet{fich89} to calculate distances. Assuming a common velocity of
$v=-45\pm6$~\kms, we find distances of $3.1\pm 0.3$~kpc or $11.1\pm0.3$~kpc.
\citet{georgelin94} identify two populations of \HII\ regions within the
Scutum-Crux arm, one at $v=-40$~\kms\ and another at $v=-50$~\kms.  While
they associate the RCW~94-95 complex with the $v=-40$~\kms\ population, our
absorption velocities indicate absorption out to nearly $v=-50$~\kms.

In order to resolve the distance ambiguity \citet{georgelin94} have
identified stars of spectral type O to B3 in the vicinity of \HII\ regions
near $l=328\arcdeg$.  They identify an O6f star, LSS~3386, in the vicinity
of RCW~94 at a distance of $2.3$~kpc.  They also identify an O7V star,
BDMW123, at a distance of 3.3~kpc near RCW~95.  While it is not clear
whether these stars are {\em the} ionizing stars for these regions, their
presence does seem to indicate a preference for the near distance of the
distance ambiguity.  We note, also, that at a distance of $3.1$~kpc RCW~94
has a physical diameter of $\sim 17$~pc, a typical diameter for an extended
\HII\ region.  Whereas at a distance of 11.1~kpc, the physical diameter
would be $\sim 62$~pc, unusually large for an \HII\ region.  The diameter at
the larger distance, as well as the nearby massive stars noted in
\citet{georgelin94}, are evidence favoring a distance of $3.1$~kpc.

The \HI\ emission morphology in the RCW~94-95 region between velocities of
$-35$~\kms\ and $-50$~\kms\ is complicated.  We have detected an \HI\ shell
surrounding RCW~94 centered at $v=-38$~\kms.  There is a ridge of \HI\
centered at $l=326\fdg3$, $b=+0\fdg8$ that lies just outside the continuum
emission contours.  This shell is shown in Fig.~\ref{fig:rcw94}, where the
greyscale is the \HI\ channel image at $v=-38$~\kms\ and the contours are
21-cm continuum emission.  The shell is apparent from $v=-35$~\kms\ to
$v=-42$~\kms. The shell itself is surrounded by a ring of decreased \HI\
emission which, although not continuous, is also centered on the \HII\
region.  The emission shell has an average diameter of $\sim 24$~pc, a
thickness of $\sim 5$~pc, and shows a brightness temperature increase from
the interior to the shell edge of about a factor of two.  The \HI\ shell
morphology closely matches the morphology of the \HII\ regions, implying
that the shell is indeed related to RCW~94.

The lack of \HI\ inside to the shell is clearly due to ionization in the
\HII\ region.  The origins of the shell are somewhat less clear.  Using the
column density integrated through the range of velocities including the
shell we estimate that the \HI\ mass of the shell is $\sim 170~{\rm
M_{\odot}}$.  That the shell extends over $\sim 7$~\kms\ suggests that it
may be expanding.  The velocity gradient at this place in the Galaxy is
$\sim 18~{\rm km~s^{-1}~kpc^{-1}}$.  Therefore a static shell with velocity
width $\Delta v \approx 7$~\kms\ would have an extent of 380~pc along the
line-of-sight.  Since it is highly unlikely that the shell extends that far,
we suggest that the velocity width is due to expansion such that $v_{exp} =
\Delta v / 2$.  Assuming $v_{exp}\sim4$ \kms, we estimate the energy
required to form this shell is on the order of $\sim 10^{51}$ ergs, which is
consistent with the amount of energy expected from stellar winds over the
lifetime of a single massive star.  Because of the low expansion velocity,
the formation energy for a shell whose expansion has stalled is comparable.

We suggest that the \HI\ shell and depression around RCW~94 are the
signatures of a molecular cloud encircling the \HII\ region.  In this case,
the \HII\ region appears to be embedded in a molecular cloud, displaying
various stages of ionization and dissociation related to the interior stars.
Interior to the $\sim 24$~pc inner shell radius the UV photons from the
stars ionize the neutral gas, producing the \HII\ region.  The stars
photo-dissociate the surrounding molecular gas, producing an \HI\ shell
which extends to a radius of $\sim 29$~pc.  The \HI\ morphology
correspondence with the continuum morphology especially supports this
hypothesis.  In particular, the region of dense \HI\ emission in the concave
portion on the right-hand side of the \HII\ region indicates that the
expansion of the photo-dissociation region (PDR) was impeded by a density
enhancement in the external medium, presumably clumps of molecular material.
The extension of the shell surrounding the compact source to the upper left,
indicates that the shape of the shell is directly related to the shape of
the \HII\ region, and that they are therefore correlated.  Comparison with
the CO images of \citet{bronfman89} indicates molecular gas at the position
and velocity of RCW~94.  Immediately exterior to the \HI\ shell we can
expect to see emission from polycyclic aromatic hydrocarbons (PAHs) at $6.2,
7.7, 8.6,$ or 11.3 \micron\ \citep{simpson99}.  Close examination of MSX
band A data does reveal an increase in emission exterior to the \HI\ shell,
which also supports the theory that this \HII\ region and its \HI\ shell are
embedded in a molecular cloud.  \citet{gaensler00a} explore the polarization
properties of this region and find that the depolarization is consistent
with being caused by an \HII\ region embedded in molecular gas with several
layers of ionization and photo-dissociation.

The \HI\ depression around RCW~94 appears to extend towards the Plane at
lower longitudes where it traces the morphology of the large bow shaped
structure, G326.96+0.03, seen at the bottom of the continuum image.  This
source is seen in the MOST images, as well as the MSX images, and therefore
appears to be a thermal source.  We measure \HI\ absorption towards the knot
of emission at $l=326\fdg95$, $b=+0\fdg02$.  Though the spectrum is rather
noisy (Fig.~\ref{fig:g327abs} left), we see a strong absorption feature at
$v=-60$~\kms, which indicates that this region may be slightly more distant
than RCW~94-95 ($d=3.9$~kpc), though still in the Scutum-Crux arm.  There is
an extended region of \HI\ emission to the left of these \HII\ regions which
is much brighter than that surrounding RCW~94-95.  The column density in
this region, over the range of channels spanning the depression
($v=-38.4~{\rm km~s^{-1}}$ to $v=-33.45~{\rm km~s^{-1}}$), is a factor of
two larger than it is surrounding the \HI\ shell.

\subsection{HI Shells}
\label{subsec:shells}
Whereas \HII\ regions and SNRs draw a connection between the \HI\ line and
continuum emission, the impact of massive stars on the ISM can also be seen
with \HI\ shells, where no continuum object exists.  These cavities survive
much longer than the radiative lifetime of a SNR or an \HII\ region,
allowing us to explore the lasting effects of massive stars on the ISM.
\HI\ shells are often detected as voids in the \HI\ with brightened
``walls'' of swept-up material.  These shells can range in size from tens of
parsecs to kiloparsecs.  The majority of the shells, especially the smaller
ones, are caused by the combined effects of stellar winds and supernovae
\citep{heiles84}.  The ultimate destruction of an \HI\ shell occurs on the
time-scale of tens of millions of years when they eventually dissipate as a
result of turbulent motions in the ISM and shear due to differential
rotation in the Galaxy.

We have detected two small shells in the SGPS Test Region
\citep{mcgriff00a}.  The first of these appears as an \HI\ void at
$l=329\fdg3$, $b=+0\fdg4$, $v=-108$~\kms, the terminal velocity for this
line of sight.  The velocity implies a kinematic distance of $7.3$~kpc.
Fig.~\ref{fig:term} is an \HI\ channel image at $v=-108$~\kms\ showing a
small shell of angular diameter is $\sim 0\fdg4$.  At a distance of
$7.3$~kpc the shell has a physical diameter of $\sim 50$~pc.  Because of its
position at the terminal velocity it is very difficult to distinguish the
front and back caps, we detect only the front cap.  It is not unusual to
detect only one cap, though.  There is only one detectable cap for a large
majority of the shells catalogued by \citet{heiles84}.  Detecting only one
cap makes it difficult to estimate an expansion velocity.  It may be that
the shell is stalled or that the structure is mostly cylindrical and
expanding in the plane of the sky.  Though we cannot measure the expansion
velocity, we interpret this structure as a stalled wind or supernova blown
shell.

The second shell is observed in the local gas at $v=-2.1$~\kms,
$l=330\fdg5$, and $b=+2\fdg12$.  This shell is shown in
Fig.~\ref{fig:local}, a channel image of the \HI\ at $v=-2.1$~\kms.  The
shell is remarkably circular with an angular diameter of $\sim 2\fdg5$.
Because of its low velocity its distance is very uncertain, we estimate $D=
350 - 500$~pc, which implies a physical radius of only $\sim 15$~pc.  Given
its small size we speculate that this shell may have been formed by an old
SNR.  There are no associated features in the continuum image.

\subsection{Supernova Remnants}
\label{subsec:snrs}

\subsubsection{SNR G327.4+0.4}
\label{subsubsec:g327.4}
\HI\ studies of supernova remnants offer a great deal of information.  \HI\ 
absorption spectra allow us to place limits on the distances to SNRs, which
lead to physical radii as well as age estimates.  Examination of related
\HI\ emission structures may help us to understand why shell-type remnant
morphology is dominated by loops and knots, in addition to diffuse emission.
The morphology of SNRs is undoubtedly related in some complicated way to the
inhomogeneities in the ISM into which they expand, as well as to non-uniform
magnetic fields.  One might expect SNR continuum emission to trace the local
ISM in such a way that bright emission may be correlated with density
enhancements exterior to the SNR shell.  In principle these ISM density
enhancements would be apparent as brightness temperature enhancements in the
neutral hydrogen at velocities similar to the systemic velocity of the
remnant.  However, in practice it has proven difficult to correlate the
continuum emission with emission structures in the \HI\ (eg. Giacani et al.\
2000\nocite{giacani00}).

We have extracted an absorption spectrum towards SNR G327.4+0.4 as shown in
Fig.~\ref{fig:g327abs}, right side.  The positive and negative wiggles in
the range $-85$ to $-95$~\kms\ are a characteristic signature of variations
in the terminal velocity, and cannot be trusted as real absorption.  There
is a strong absorption line centered at $v=-48$~\kms, and a weaker, noisy
line at about $-70$~\kms.  There is no absorption corresponding to the
emission peak at $-80$~\kms.  We therefore adopt$V_{\rm L}=-70$~\kms\ and
$V_{\rm U}=-80$~\kms.  These velocities indicate a distance of $4.3\pm0.5$~
kpc.  This places the remnant on the far side of the Scutum-Crux Arm.

SNR G327.4+04 is a multi-arc shell-type SNR.  The continuum emission has a
bright, sharply bounded rim to the lower left, while the emission on the
upper right side is much more diffuse.  If the limb brightening observed to
the lower left is a consequence of the shock impacting a density
enhancement, we might expect to see an \HI\ cloud exterior to the shell at
the systemic velocity of the remnant.  This is confirmed in the \HI\ channel
images at $v\approx -70$~\kms, where we see a ridge of emission just exterior
to the continuum contours.  Fig.~\ref{fig:snr327.4} is an average of two
velocity channels centered at $v=-70$~\kms\ with continuum contours
overlaid.  There is an increase in \HI\ density just exterior to the
brightest portion of the SNR.  In addition, the \HI\ to the upper right,
exterior to the less bright edge of the remnant, is much more diffuse.  In
this case it does appear that the continuum morphology is related to the
surrounding \HI.

SNR G327.4+0.4 has an angular diameter of $\sim 15\farcm3$, yielding a
physical radius of $10.5 \pm 0.6$~pc at $4.3$~kpc.  To reinforce the point
that this remnant must be at the near distance, we note that at $10.0$~kpc
the physical radius of the SNR would be an unusually large 22~pc. The radius
allows us to estimate some fundamental parameters for the SNR.  If the
density of the medium into which the SNR expands is given by $n_o$ (in ${\rm
cm ^{-3}}$), the mass swept up by the SNR is $\sim 120 n_o~{\rm M_{\odot}}$.
As this mass is only a few times the presumed mass of the progenitor, we
believe that the SNR is undergoing adiabatic expansion, but that it has only
recently left the free expansion phase of evolution.  Using the standard
assumption that the SNR is in the Sedov-Taylor phase, we estimate the age
$t_{\rm SNR} = (5.3\pm 0.8) (n_o/E_{51})^{1/2} \times 10^3$~yr, where
$E_{51}$ is the input energy of the supernova explosion in units of
$10^{51}$~ergs.  If we assume typical values of $n_o=0.2$ and $E_{51}=1$
\citep{frail94}, we find $t_{\rm SNR} = (2.4 \pm 0.3) \times 10^3$~yr.

\subsubsection{SNR G330.2+1.0}
\label{subsubsec:g330.2}
We obtain an absorption spectrum towards SNR G330.2+1.0 which shows
absorption out to $v=-80\pm 6$~\kms, indicating a minimum distance of
$d=4.9\pm 0.3$~kpc for the supernova remnant.  As noted in \S\ref{sec:cont},
this SNR has no clearly defined shell.  There is continuum emission
surrounding the brightened center.  It is not clear how much of that
emission is associated with the SNR.  Fig.~\ref{fig:snr330.2} is an \HI\
image at $v=-80$~\kms overlaid with 21-cm continuum contours of SNR
G330.2+1.0.  At the low longitude end of G330.2+1.0 the lowest continuum
contour extends in an arc away from the SNR center.  Spatially offset from
this edge the \HI\ emission follows the same arc.  Similarly, the continuum
contour closest to the Plane is bounded by \HI\ emission which traces the
contour.  The morphological similarities between the lowest continuum
contour and the \HI\ emission to the right of the remnant suggests a
possible correlation between the two.  The overlap of \HI\ from two
distances at this velocity, as well as the small angular size of the SNR
make it difficult to confirm whether the apparent \HI-continuum correlation
is real.

\section{Conclusions}
\label{sec:concl}
We have presented \HI\ and $\lambda 21$-cm data from the SGPS Test Region
($325\fdg5 \leq l \leq 333\fdg5$; $-0\fdg5 \leq b \leq 3\fdg5$), which are
representative of the full Survey.  These results highlight the interesting
effects of massive stars on the ISM.  The SGPS is ideal for studying the
structure and dynamics of the \HI\ in the inner Galaxy as it is sensitive to
a large range of angular scales ($2\arcmin \lesssim \theta \lesssim
2\arcdeg$).  In the Test Region we have explored the \HI\ associated with
three products of massive star life and death: \HII\ regions, SNRs, and \HI\
shells.  Using \HI\ absorption for systemic velocities and corresponding
kinematic distances of \HII\ regions and SNRs we are able to create a three
dimensional picture of the distribution of the continuum sources in this
region of the Galaxy.

We have highlighted several interesting \HI\ and $\lambda 21$-cm continuum
emission features from the Test Region.  The features in the continuum image
include extended emission structures and a possible new SNR G328.6-0.0.
Comparing the \HI\ and continuum, we found an \HI\ shell around the \HII\
region RCW~94 which indicates that the \HII\ region is embedded in a
molecular cloud.  In this case we see the reciprocal effects of massive
stars and the surrounding \HI\ during the stellar lifetime.  The continuum
emission morphology of the \HII\ region closely matches the morphology of
the surrounding \HI.  We use \HI\ absorption towards the SNRs G327.4+0.4 and
G330.2+1.0 to determine kinematic distances of 4.3 and 4.9~kpc,
respectively.  \HI\ at the systemic velocity of these remnants shows
morphological similarities to the continuum emission.  In particular,
density enhancements were found exterior to regions of continuum
limb-brightening for G327.4+0.4.  We also found two small \HI\ shells with
no counterparts in continuum emission.  We use the sizes and lack of
detectable expansion velocity to interpret these structures as stalled
supernova or wind blown shells which are older than the radiative lifetimes
of either \HII\ regions or SNRs.

Deciphering \HI\ structure has always been challenging, but the recent
availability of high resolution Galactic surveys such as the SGPS has
improved the situation dramatically.  Much of the inner Galaxy is completely
filled with a variety of \HI\ structures including shells, worms, sheets,
and filaments.  Though it is extremely difficult to determine the origins of
many of the structures using \HI\ emission data alone, combination with \HI\ 
absorption and radio continuum emission measurements enables us to determine
a three-dimensional, dynamical picture of the ISM.

\acknowledgements We thank Veta Avedisova for supplying us with her
extensive catalogue of star formation regions.  This research has made
use of the CDS SIMBAD database.  JMD and NMM-G acknowledge support of
NSF grant AST-9732695 to the University of Minnesota.  NMM-G is
supported by NASA Graduate Student Researchers Program (GSRP)
Fellowship NGT 5-50250.  BMG acknowledges the support of NASA through
Hubble fellowship grant HST-HF-01107.01-A awarded by STScI, which is
operated by AURA Inc. for NASA under contract NAS 5-26555.

\vfil
\eject 
\begin{figure}
\begin{center}
  \figcaption[f1.eps]{21-cm continuum image of the SGPS Test Region from the
    ATCA with the 190 pointing centers marked.  The greyscale is linear and
    runs from $-0.03~{\rm mJy~beam^{-1}}$ to $0.1~{\rm mJy~beam^{-1}}$ as
    shown in the wedge at the right.  The image has an rms noise of
    $5.5~{\rm mJy~beam^{-1}}$.  The beam is $124\farcs9 \times 107\farcs5$
    and is displayed in the upper left of the image.
\label{fig:centers}}
\end{center}
\end{figure}

\begin{figure}
\figcaption[f2a.eps]{Channel images from the combined Parkes and ATCA
continuum subtracted \HI\ line cube of the Test Region.  Every fourth
channel is displayed for a velocity separation of 3.2~\kms.  The greyscale
is linear from 0~K to 115~K and shown in the wedge at the right.  The beam
size is $124\farcs9 \times 107\farcs5$ and the rms noise is $\sim 2.3$~K.
\label{fig:chans1}}
\end{figure}
\setcounter{figure}{1}
\begin{figure}
\figcaption[f2b.eps]{Continued
\label{fig:chans2}}
\end{figure}
\setcounter{figure}{1}
\begin{figure}
\figcaption[f2c.eps]{Continued
\label{fig:chans3}}
\end{figure}
\setcounter{figure}{1}
\begin{figure}
\figcaption[f2d.eps]{Continued
\label{fig:chans4}}
\end{figure}

\setcounter{figure}{2}
\begin{figure}
\begin{center}
  \figcaption[f3.eps]{Continuum image of the 21-cm emission in the SGPS Test
  Region.  The image includes both ATCA and Parkes data.  The greyscale is
  linear and runs from 0.5 to $1.1~{\rm Jy~beam^{-1}}$ as shown in wedge at
  the right.  The rms noise is $\sim 7~{\rm mJy~beam^{-1}}$.  The beam is
  shown in the upper left corner and is $124\farcs9 \times 107\farcs5$.
\label{fig:21cont}}
\end{center}
\end{figure}

\begin{figure}
\begin{center}
 \includegraphics[scale=0.7,angle=-90]{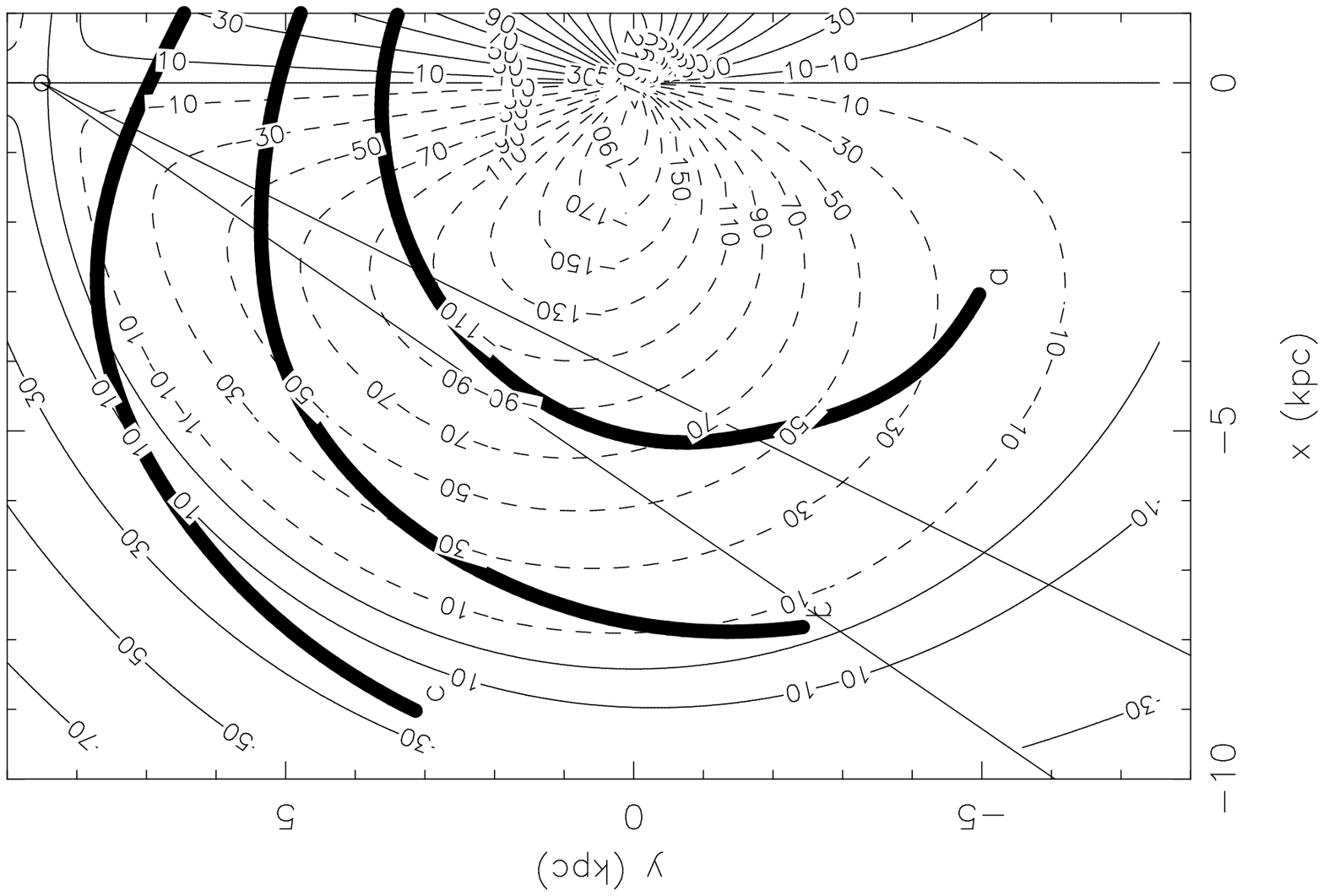}
  \figcaption[f4.eps]{Theoretical isovelocity contours calculated from the
  rotation curve of \citet{fich89} and overlaid on the spiral pattern of the
  Galaxy from \citet{taylor93}.  The coordinates $x$ and $y$ are with
  respect to the Galactic center.  The heavy black lines mark the
  approximate positions of the spiral arms.  The spiral arms are labeled as
  {\bf a}: Norma; {\bf b}: Scutum-Crux; and {\bf c}: Sagittarius-Carina.
  Contours are labeled with their LSR velocities, where dashed contours
  denote negative velocities and solid contours denote positive velocities.
  The two thin lines forming a wedge mark the SGPS Test Region line of
  sight.  The line of sight covers the theoretical velocity range $-120~{\rm
  km~s^{-1}} \leq v \leq 80~{\rm km~s^{-1}}$.  The approximate positions of
  the spiral arms in velocity space can be discerned from the velocity
  contours that they cross.
\label{fig:diag}}
\end{center}
\end{figure}

\begin{figure}
\begin{center}
\figcaption[f5.eps]{843 MHz continuum image of the SGPS Test Region from the
  MOST Galactic Plane Survey \citep{green99}.  Prominent \HII\ regions and
  SNRs are labeled.  The greyscale is linear and runs from $-30~{\rm
  mJy~beam^{-1}}$ to $60~{\rm mJy~beam^{-1}}$, as shown in the wedge at the
  left.  The beamsize for the MOST survey is $43\farcs0 \times 51\farcs9$
  and the rms noise is $\sim 2.2~{\rm mJy~beam^{-1}}$.
\label{fig:most}}
\end{center}
\end{figure}

\begin{figure}
\begin{center}
\epsscale{0.5}
\figcaption[f6.eps]{Images of a possible new SNR G328.6-0.0.  To the left,
  image a, is 843 MHz continuum emission from the MOST Galactic Plane Survey
  \citep{green99}.  The image at the right, image b, is 1420 MHz continuum
  emission from the SGPS.  The rms noise of the MOST image is $\sim 2.2~{\rm
  mJy~beam^{-1}}$ and the beamsize is $43\farcs0 \times 51\farcs9$.  The rms
  noise of the SGPS image is $\sim 7~{\rm mJy~beam^{-1}}$ and the beamsize
  is $124\farcs9 \times 107\farcs5$.  The possible SNR is the large loop in
  the center of the image.
\label{fig:newsnr}} 
\end{center}
\end{figure}

\begin{figure}
\begin{center}
\epsscale{0.4} 
\plotone{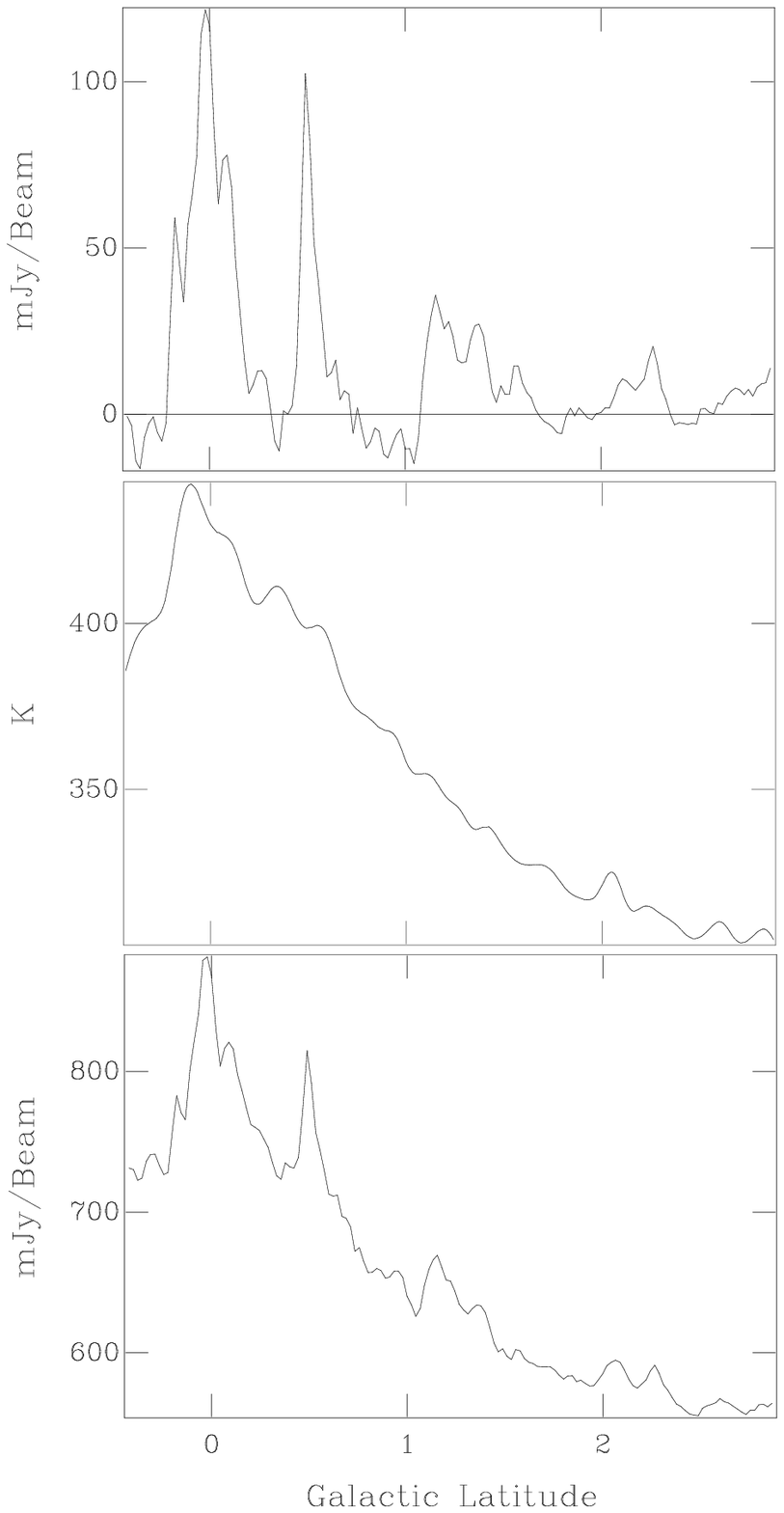} 
\figcaption[f7.eps]{Slices across the SGPS Test Region continuum image at
$l=328\fdg7$ using only the ATCA data (top), only the Parkes data (center),
and the combined Parkes and ATCA data (bottom).  Individual ources are
resolved in the ATCA slice, whereas the Parkes data detects the Galactic
background.  The combined data show individual sources superimposed on the
Galactic background.
\label{fig:contslice}}
\end{center}
\end{figure}

\begin{figure}
\epsscale{0.9} 
\plotone{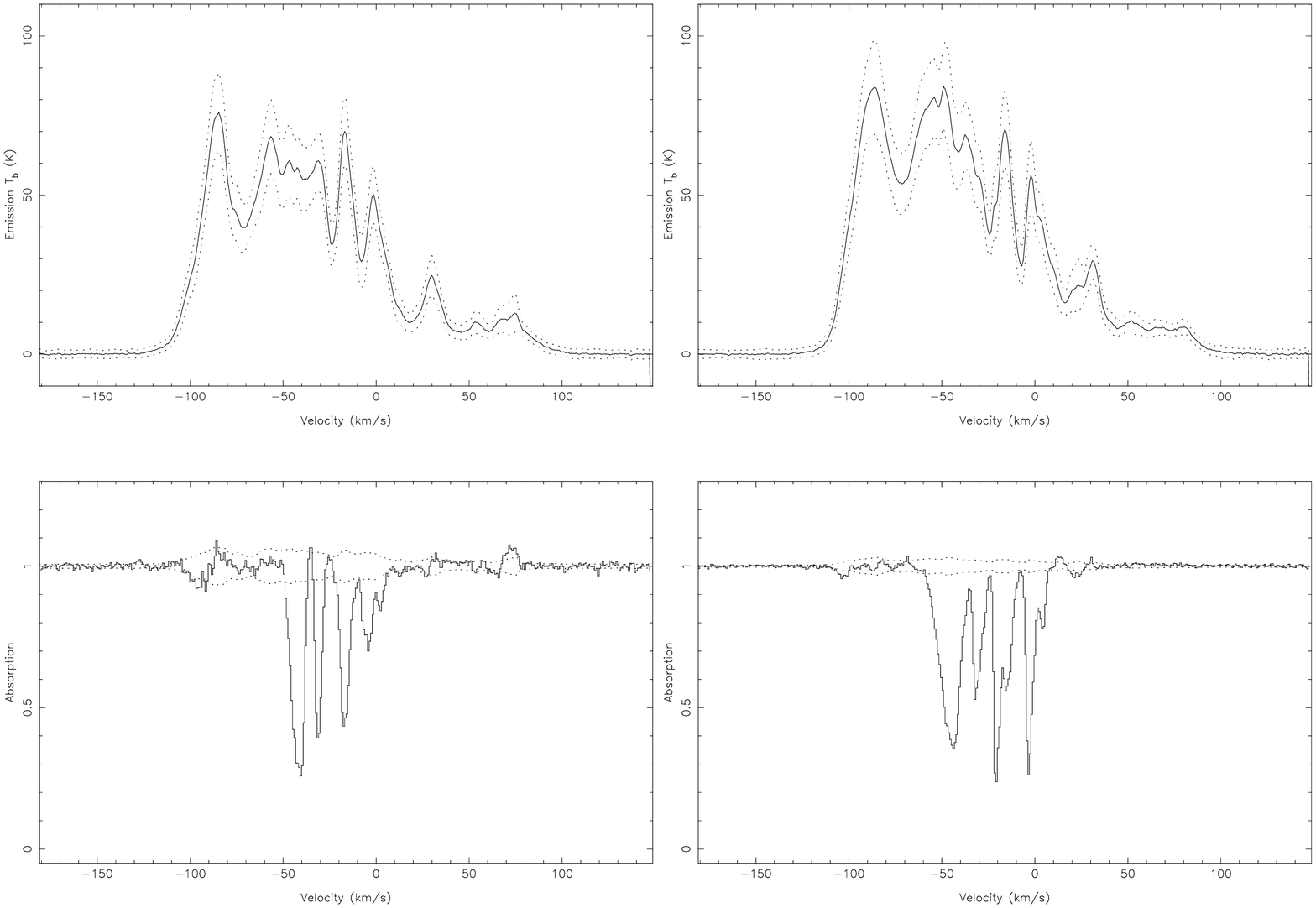} 
\figcaption[f8.eps]{\HI\ emission (top) and absorption (bottom) spectra
taken towards RCW~94 (left) and G326.65+0.59 (right).  The top panels are
interpolated emission around the source.  The dashed lines mark the
calculated $1\sigma$ error envelope of the interpolated emission.  The
bottom panels plot $e^{-\tau}$ and associated $1\sigma$ errors towards the
\HII\ regions.
\label{fig:rcw94_abs}}
\end{figure}

\begin{figure}
\epsscale{0.9} 
\plotone{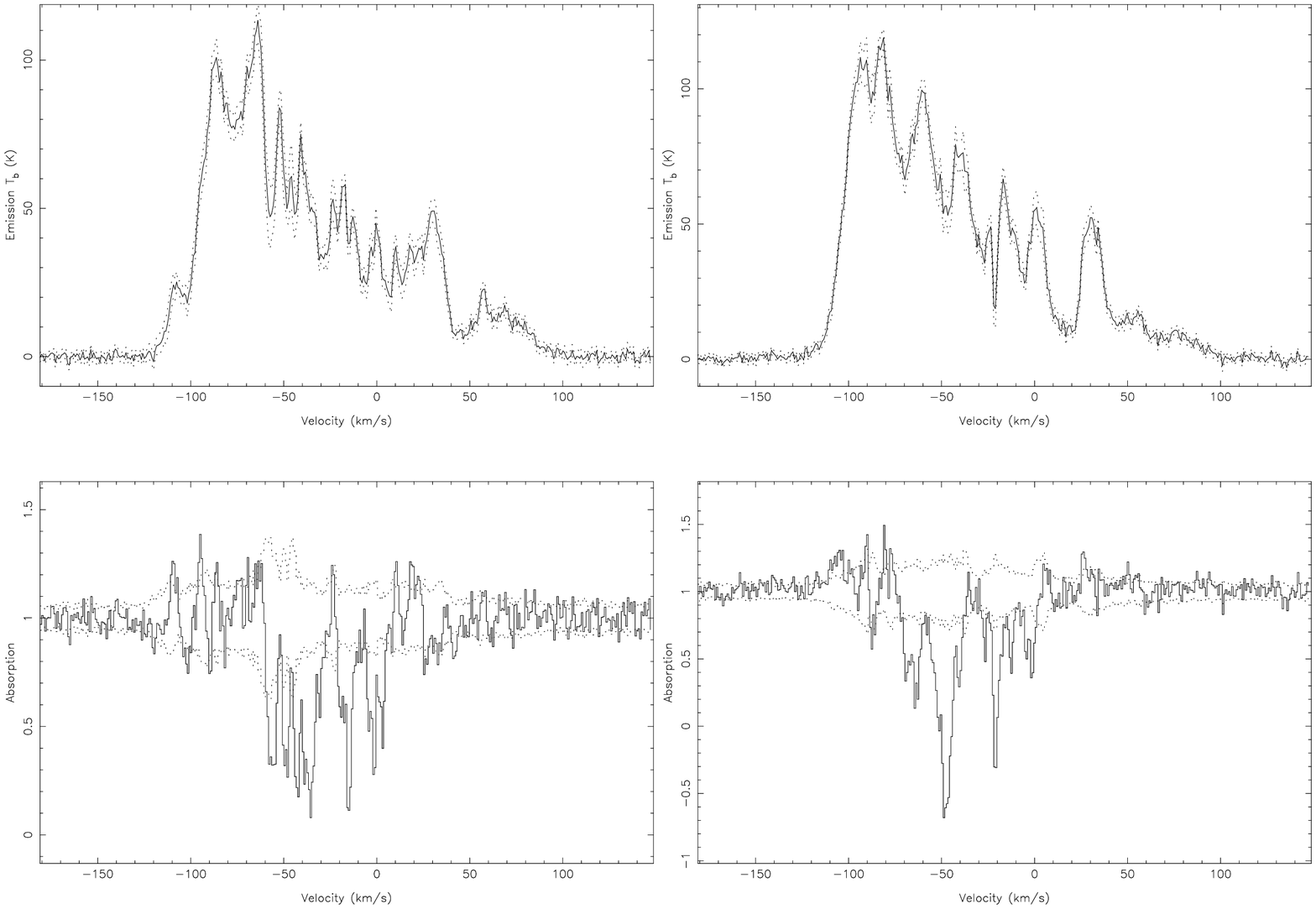} 
\figcaption[f9.eps]{\HI\ emission and absorption towards G326.96+0.03 (left)
and towards SNR G327.4+0.4 (right).  The panels are the same as
Fig.~\ref{fig:rcw94_abs}.
\label{fig:g327abs}}
\end{figure}

\begin{figure}
\begin{center}
  \includegraphics[scale=0.4, angle=-90]{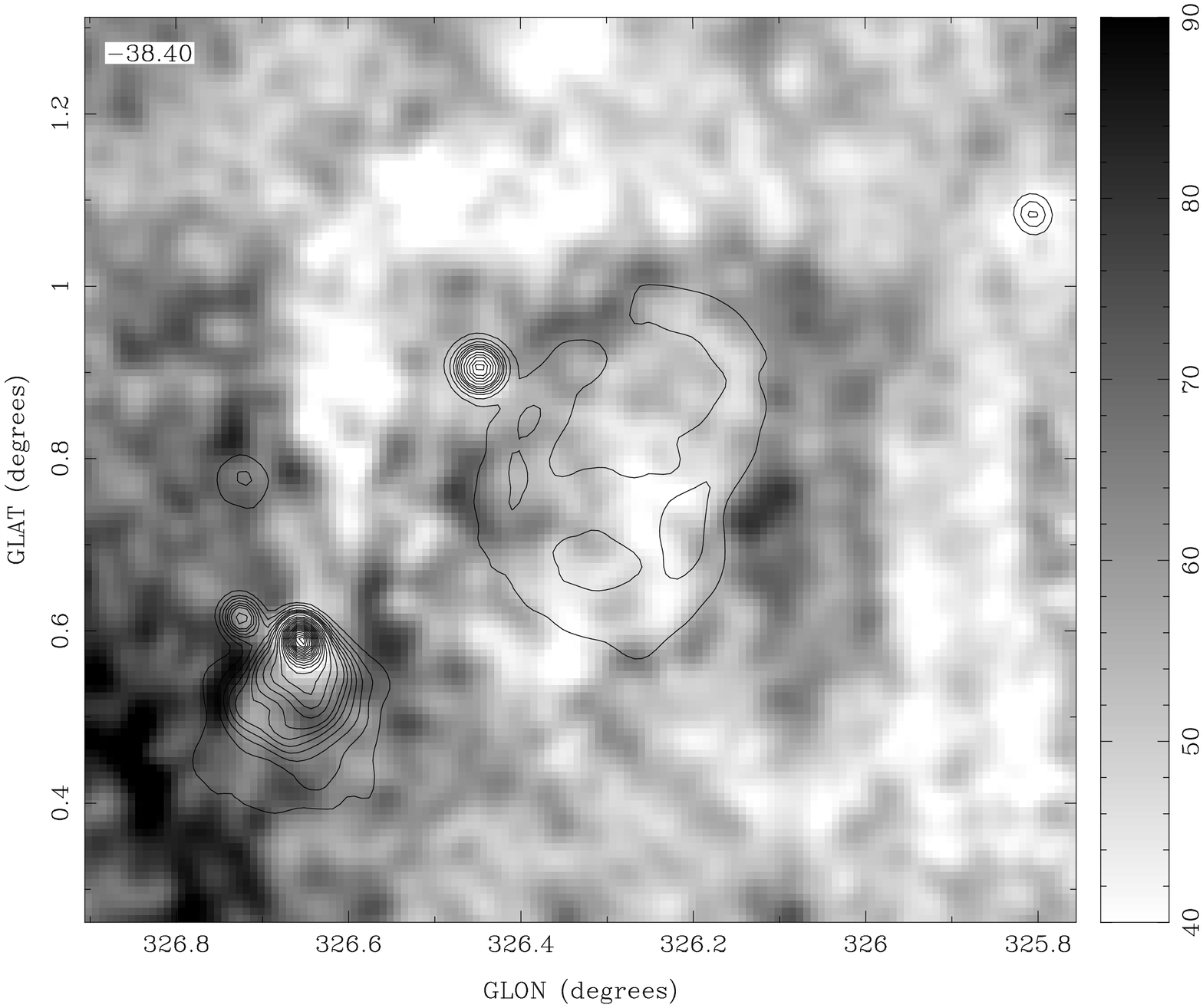}
  \figcaption[f10.eps]{Greyscale \HI\ channel image at $v=-38$~\kms\ of the
  region surrounding the RCW~94-95 \HII\ complex; 21-cm continuum contours
  are overlaid.  The greyscale is linear and runs from 40 to 90 K as shown
  in the wedge on the right.  The continuum contours are at $0.2~{\rm
  Jy~beam^{-1}}$ intervals from $0.6~{\rm Jy~beam^{-1}}$ to $10~{\rm
  Jy~beam^{-1}}$.  The beam size is $124\farcs9 \times 107\farcs5$ and the
  rms noise is $\sim 2.3$~K in the line and $\sim 7~{\rm mJy~beam^{-1}}$ in
  the continuum.  The small \HI\ shell, centered at $l=326\fdg3$,
  $b=+0\fdg8$ is identified by the ring of \HI\ emission around the \HII\
  region contours.
\label{fig:rcw94}}
\end{center}
\end{figure}

\begin{figure}
\begin{center}
  \figcaption[f11.eps]{Greyscale image of an \HI\ channel image at
  $v=-108$~\kms\ showing an apparent \HI\ shell in the ISM at the terminal
  velocity.  The greyscale is linear from 0~K to 80~K, as shown in the wedge
  on the right.  The beam size is $124\farcs9 \times 107\farcs5$ and the rms
  noise is $\sim 2.3$~K.  The small shell, of diameter $\sim 0\fdg4$, is
  located at $l=329\fdg3$, $b=+0\fdg4$ and is identified by the bright ring
  of emission surrounding the \HI\ void.
\label{fig:term}}
\end{center}
\end{figure}

\begin{figure}
\begin{center}
  \figcaption[f12.eps]{Greyscale \HI\ channel image at $v=-2.12$~\kms\
  showing an apparent \HI\ shell in the local ISM at $l=330\fdg5$,
  $b=+2\fdg2$.  The greyscale is logarithmic to emphasize the shell walls.
  The brightness temperature scale in Kelvins (35~K to 90~K) is displayed in
  the wedge on the right.  The beam size is $124\farcs9 \times 107\farcs5$
  and the rms noise is $\sim 2.3$~K.
\label{fig:local}}
\end{center}
\end{figure}
\begin{figure}
\begin{center}
  \includegraphics[scale=0.4,angle=-90]{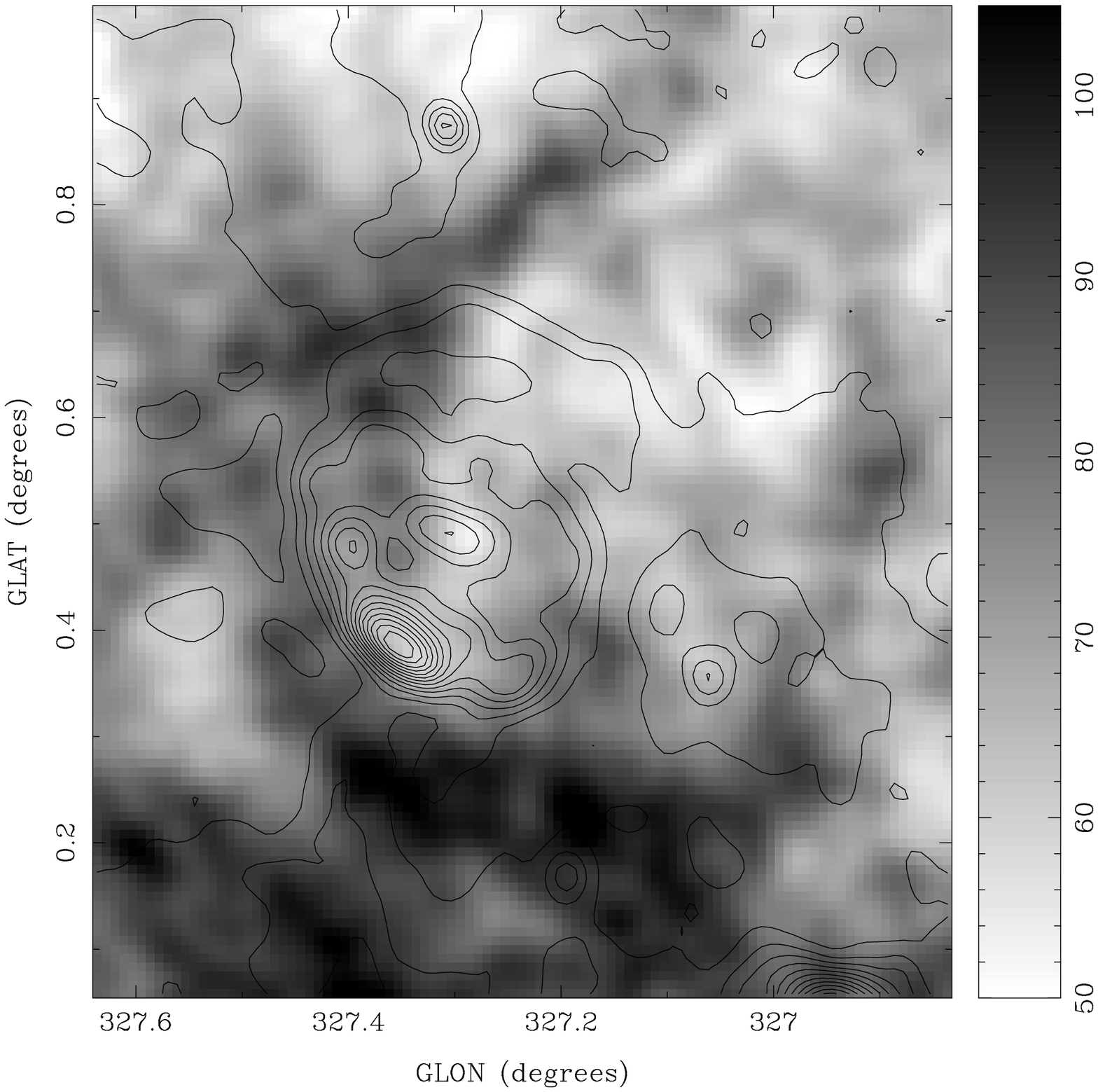} 
\figcaption[f13.eps]{SNR G327.7+0.4.  Greyscale image of an average of two
\HI\ velocity channels centered at $v=-70$~\kms.  The greyscale is linear
and runs from 50 to 105 K, as shown in the wedge at the right.  The contours
show the 21-cm continuum emission from SNR G327.7+0.4 at intervals of
$60~{\rm mJy~beam^{-1}}$ from $300~{\rm mJy~beam^{-1}}$ to $2~{\rm
Jy~beam^{-1}}$. The beam size is $124\farcs9 \times 107\farcs5$ and the rms
noise is $\sim 2.3$~K in the line and $\sim 7~{\rm mJy~beam^{-1}}$ in the
continuum.
\label{fig:snr327.4}}
\end{center}
\end{figure}

\begin{figure}
\begin{center}
  \includegraphics[scale=0.5,angle=-90]{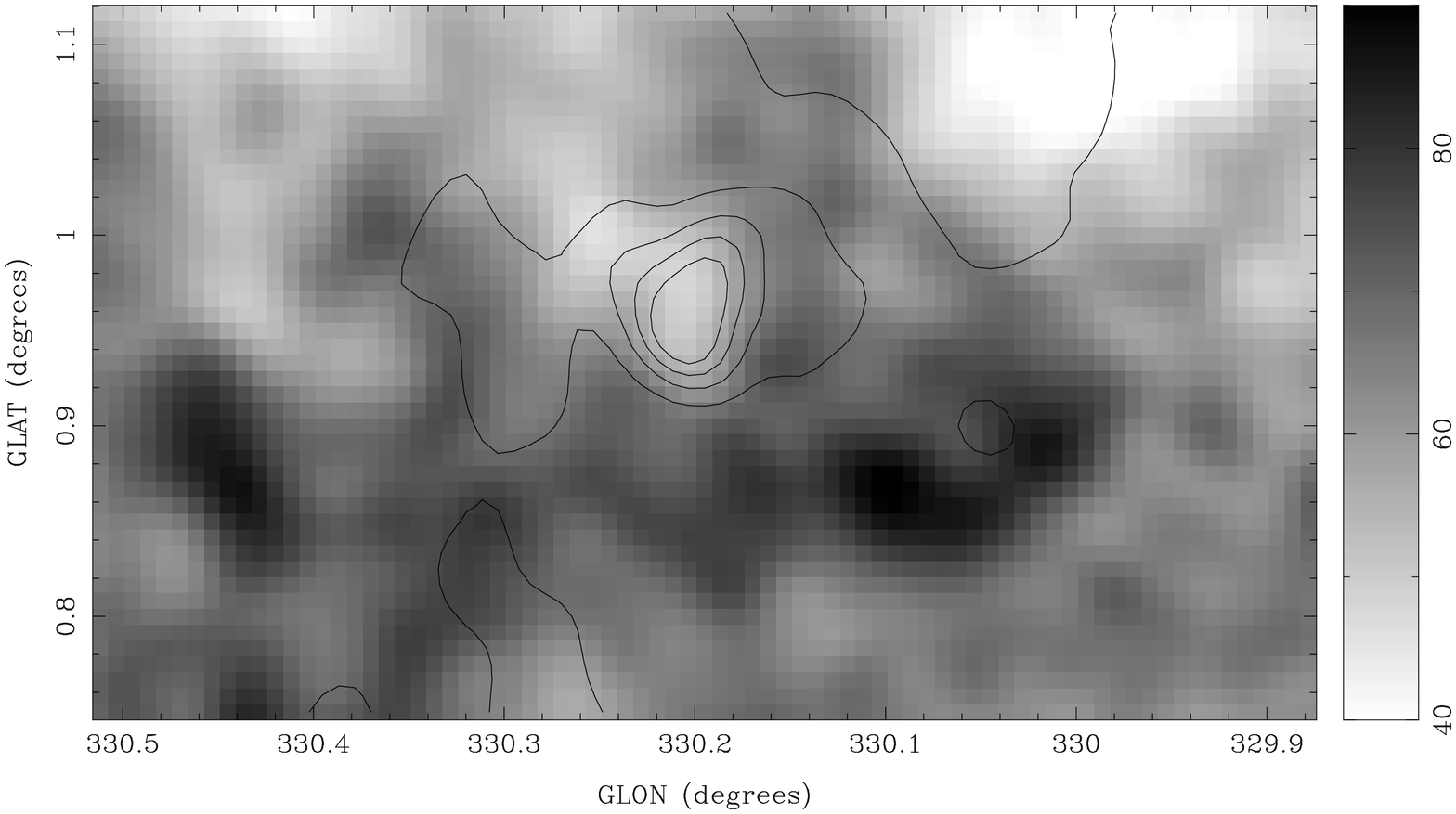} 
\figcaption[f14.eps]{SNR G330.2+1.0.  Greyscale image of an average of two
  \HI\ velocity channels centered at $v=-80$~\kms.  The greyscale is linear
  and runs from 40 to 90 K, as shown in the wedge at the right.  The
  contours show the 21-cm continuum emission from SNR G330.2+1.0 from
  $210~{\rm mJy~beam^{-1}}$ to $2.1~{\rm Jy~beam^{-1}}$ at intervals of
  $105~{\rm mJy~beam^{-1}}$.  The beam size is $124\farcs9 \times
  107\farcs5$ and the rms noise in the line is $\sim 2.3$~K and $\sim 7~{\rm
  mJy~beam^{-1}}$ in the continuum.  The \HI\ emission roughly traces the
  outermost continuum contour.
\label{fig:snr330.2}}
\end{center}
\end{figure}

\clearpage
\begin{deluxetable}{lll}
\tabletypesize{\scriptsize}
\tablecaption{Table of observing dates and telescope or array configurations
  for the Test Region. 
\label{tab:obs}}
\tablewidth{0pt}
\tablehead{
\colhead{Date} & \colhead{Array} & \colhead{UT} 
}
\startdata
1997 Apr 21 &  0.375 & 09:22 - 22:14\\
1997 Apr 22 &  0.375 & 08:00 - 22:05\\
1997 Apr 23 &  0.375 & 12:11 - 22:03 \\
1997 Aug 11 &  0.75B & 05:33 - 16:06\\
1997 Aug 13 &  0.75B & 08:33 - 15:51\\
1997 Oct 25 &  0.75C & 20:53 - 11:12\\
1998 Mar 30 &  0.375 & 10:00 - 20:40\\
1998 Apr 20 &  0.75A & 08:06 - 20:59\\
1998 Apr 21 &  0.75A & 07:44 - 21:02\\
\tableline
1998 Dec 14-15 & Parkes & 20:16 - 05:21\\
1998 Dec 15-16 & Parkes & 20:22 - 06:33\\
1998 Dec 16 & Parkes & 14:56 - 22:30\\
\enddata
\end{deluxetable}

\clearpage
\begin{deluxetable}{llrcccccc}
\tabletypesize{\scriptsize}
\tablecaption{Table of \HI\ absorption velocities and kinematic distances
for extended continuum sources. 
\label{tab:abs}}
\tablewidth{0pt}
\tablehead{
& & & \colhead{\HI\ } & \colhead{\HI\ } & \colhead{H109$\alpha$ \&
H110$\alpha$\tablenotemark{b}} & & & \\
\colhead{Source} & \colhead{$l$} & \colhead{$b$} &
\colhead{$V_{L}$\tablenotemark{a}} &\colhead{$V_{U}$} & \colhead{$V$}
& \colhead{$D_L$\tablenotemark{c}} & \colhead{$D_U$\tablenotemark{c}}
& \colhead{$R_{gal}$\tablenotemark{d}}\\
& \colhead{(deg)} & \colhead{(deg)} & \colhead{(\kms)} &
\colhead{(\kms)} & \colhead{(\kms)} & \colhead{(kpc)} &\colhead{(kpc)}
& \colhead{(kpc)}
}
\startdata
RCW~94 & 326.45 & $+0.91$ & $-42$ & $-48$ & $-45$ & 2.9 & 3.3 & 6.2\\
G326.65+0.59 & 326.66 & $+0.59$ & $-47$ & $-60$ & $-44$ & 3.2 & 3.9 & 6.0\\
G326.96+0.03 & 326.95 & $+0.02$ & $-57$ & $-62$ & $-64$ & 3.7 & 4.0  & 5.6\\
SNR G327.4+0.4 (Kes 27)& 327.34 & $+0.40$ & $-67$ & $-82$ & \nodata & 4.3 &
5.4 & 5.2\\
G327.99-0.09 & 327.99  & $-0.09$ & $-52$ & $-68$ & $-45$ & 3.5 & 4.3 & 5.7\\
G328.31+0.45 & 328.30 & $+0.44$ & $-96$ & $-101$ & $-97$ & 6.0 & 6.5 & 4.6\\
SNR G328.4+0.2 (MSH 15-57)& 328.42 & $+0.22$ & $+28$ & \nodata & \nodata &
17.4  & \nodata & 11.1\\
G329.35+0.14 & 329.34 & $+0.14$ & $-103$ & $-109$ & $-107$ & 6.4 & 7.3 & 4.4\\
G329.49+0.21 & 329.47 & $+0.21$ & $-100$ & $-109$ & $-102$ & 6.1 & 7.3 & 4.5\\  
SNR G330.2+1.0 & 330.21 & $+0.97$ & $-80$ & \nodata & \nodata &
4.9(9.9)\tablenotemark{e}  & \nodata & 4.9\\
G331.03-0.15 & 331.05 & $-0.16$ & $-95$ & $-100$ & $-89$ & 5.5 & 5.9 & 4.5\\
G331.26-0.19 & 331.27 & $-0.19$ & $-89$ & $-100$ & $-85$ & 5.3 & 5.9 & 4.5\\  
G331.52-0.07 & 331.52 & $-0.08$ & $-92$ & $-102$ & $-89$ & 5.5 & 6.0 & 4.4\\
\enddata
\tablenotetext{a}{All velocities all quoted with respect to the local
standard of rest (LSR).}
\tablenotetext{b}{From \citet{caswell87}.}
\tablenotetext{c}{Distances are derived using a standard rotation
curve as in Fich, Blitz \& Stark (1989), assuming IAU standard parameters of
$\Theta_o=220$~\kms\ and $R_o=8.5$~kpc.}
\tablenotetext{d} {$R_{gal}$ values are calculated at the lower velocity
limit.}
\tablenotetext{e}{Values in parentheses are the more distant for velocities
interior to the solar circle.}
\end{deluxetable}

\end{document}